\documentclass[11pt]{article}
\usepackage{graphicx}
\def \d {{\rm d}}
\def \o {^\circ}

\begin{document}

\title{Generalised Kundt waves and their physical interpretation} 

\author{J. B. Griffiths$^1$\thanks{E--mail: {\tt J.B.Griffiths@Lboro.ac.uk}}, \ 
P. Docherty$^1$ \ and 
J. Podolsk\'y$^2$\thanks{E--mail: {\tt Podolsky@mbox.troja.mff.cuni.cz}} \\ \\ 
$^1$ Department of Mathematical Sciences, Loughborough University, \\ 
Loughborough, Leics. LE11 3TU, U.K. \\ \\
$^2$ Institute of Theoretical Physics, Charles University in Prague,\\
V Hole\v{s}ovi\v{c}k\'ach 2, 18000 Prague 8, Czech Republic.\\
\\}

\date{\today}
\maketitle

\begin{abstract}
\noindent
We present the complete family of space-times with a non-expanding, shear-free,
twist-free, geodesic principal null congruence (Kundt waves) that are of
algebraic type~III and for which the cosmological constant ($\Lambda_c$) is
non-zero. The possible presence of an aligned pure radiation field is also
assumed. These space-times generalise the known vacuum solutions of type~N with
arbitrary $\Lambda_c$ and type~III with $\Lambda_c=0$. It is shown that there are
two, one and three distinct classes of solutions when $\Lambda_c$ is respectively
zero, positive and negative. The wave surfaces are plane, spherical or
hyperboloidal in Minkowski, de~Sitter or anti-de~Sitter backgrounds respectively,
and the structure of the family of wave surfaces in the background space-time is
described. The weak singularities which occur in these space-times are
interpreted in terms of envelopes of the wave surfaces. 
\end{abstract}

\section{Introduction}

The families of plane and {\sl pp}-waves are among the best known exact solutions
of Einstein's equations. They describe plane-fronted waves with parallel rays
which propagate in a Minkowski background. A more general family of plane-fronted
waves is also well known \cite{KSMH80} and described as Kundt's class. This is
characterised by the fact that it possesses a repeated principal null geodesic
congruence which is non-expanding and has zero twist and shear. In the original
papers of Kundt \cite{Kundt61}--\cite{Kundt62} it was shown that, for zero
cosmological constant but permitting an aligned pure radiation field, these
space-times are of algebraic types III, N or O. Together, these are characterised
as the only radiation fields which possess plane wave surfaces. The type~D
space-times which satisfy the above criteria, and some type~II space-times, are
also known (see \cite{KSMH80} and references cited therein).

The introduction of a non-zero cosmological constant ($\Lambda_c$) to the above
criteria, however, is not so easily achieved as it is for the equivalent
(Robinson--Trautman) class of expanding waves. Nevertheless, families of
non-expanding type~N vacuum space-times with $\Lambda_c\ne0$ were found by
Garc\'{\i}a D\'{\i}az and Pleba\'nski \cite{GarPle81} and analysed further by
Ozsv\'ath, Robinson and R\'ozga \cite{OzRoRo85} and by Bi\v{c}\'ak and Podolsk\'y
\cite{BicPod99a}--\cite{BicPod99b}. Also, Podolsk\'y and Ortaggio \cite{PodOrt02}
have described a large family of explicit Kundt type~II and~N solutions with
$\Lambda_c\ne0$ and interpreted them as gravitational waves in various type~D
and~O universes. However, no type~III solutions with a non-zero cosmological
constant have been studied. One purpose of the present paper is to fill this gap.
In fact, in doing so, we also find a new family of type N solutions for which the
cosmological constant is necessarily non-zero. The main purpose of the paper,
however, is to present a comprehensive derivation of all type~III solutions
within this class with a non-zero cosmological constant and an aligned pure
radiation field, and to describe some aspects of their physical interpretation at
least in the weak field limit.

\section{The line element for Kundt's class}

To derive all space-times of this class, we will employ the Newman--Penrose (NP)
formalism \cite{NewPen62} and the notations associated with it. We start with the
assumption that the space-time is of algebraic type~III, and choose the tetrad
vector $\ell^\mu$ to be everywhere tangent to the repeated principal null
congruence. This immediately implies that \ $\Psi_0=\Psi_1=\Psi_2=0$. \ In
general, we will also include a cosmological constant using the NP notation \
$\Lambda={1\over6}\Lambda_c$ \ and an aligned pure radiation field such that
$R_{\mu\nu}=-2\Phi_{22}\ell_\mu\ell_\nu$.

With the above assumptions, the repeated principal null congruence is necessarily
geodesic and shear-free so that $\kappa=\sigma=0$. To obtain Kundt's class of
solutions, we make the additional assumption that the repeated principal null
congruence is also expansion-free and twist-free, so that $\rho=0$. We are free
to choose an affine parameter $r$ along the congruence and, since $\ell_\mu$ must
be proportional to a gradient, we use the tetrad freedom to set $\ell_\mu$ equal
to a gradient. Using also a freedom in another tetrad vector, this produces the
further constraints on the spin coefficients \ $\epsilon=0$ \ and \
$\bar\alpha+\beta=\tau$.

We adopt coordinates which are labelled such that \ $r=x^1$, \ 
$\zeta={1\over\sqrt2}(x^2+ix^3)$ \ and \ $u=x^4$ \ and such that \ 
$\ell_\mu=\delta_\mu^4$ \ and \ $\ell^\mu=\delta_1^\mu$. \ We then use the
available coordinate and tetrad freedoms to put \ $\pi=-\bar\tau$ \ and \
$\lambda=0$ \ and to simplify the tetrad to the form 
 \begin{equation}
 \begin{array}{lccl}
 \ell^\mu =\delta_1^\mu 
 &&\qquad& D= {\textstyle {\partial\over\partial r} } \\[8pt] 
 m^\mu = -P\bar W\,\delta_1^\mu 
+{\textstyle{1\over\sqrt2}}P\,(\delta_2^\mu+i\delta_3^\mu) 
 &&& \delta= {\textstyle P\left({\partial\over\partial\bar\zeta}
-\bar W\,{\partial\over\partial r}\right) } \\[8pt] 
 n^\mu = -H\,\delta_1^\mu +\delta_4^\mu 
 &&& \Delta = {\textstyle {\partial\over\partial u} 
-H\,{\partial\over\partial r}} 
 \end{array}
 \label{tetrad}
 \end{equation}
 where $P$ and $H$ are real functions, $W$ is a complex function, and  \
$P_{,r}=0$. \ The line element then has the form 
 \begin{equation}
 \d s^2 =2\d u(\d r +H\d u +W\d\zeta +\bar W\d\bar\zeta) 
-2P^{-2}\d\zeta\d\bar\zeta.  
 \label{metric}
 \end{equation}
 (The details to this point are given in \cite{KSMH80}. The Ricci identities and
the metric equations are given in the appendix.)

We now note that the Ricci identity $D\tau=0$ implies that \ $\tau=\tau\o$, \ in
which we are using the standard notation that a degree sign indicates any function
(except $P$) which is independent of~$r$. One of the metric equations can then
immediately be integrated to give 
 \begin{equation}
 W={2\bar\tau\o\over P}\,r +W\o, 
 \label{omega}
 \end{equation} 
 where $W\o$ is arbitrary. Another, together with one of the Ricci
identities, implies that
 \begin{equation}
 H=-(\tau\o\bar\tau\o+\Lambda)r^2+{2G\o}r+H\o 
 \label{H} 
 \end{equation}
 where $G\o$ and $H\o$ are real functions. It can also be shown that \ 
$\beta={1\over2}(\tau\o-P_{,\bar\zeta})$.

\section{Wave surfaces}

For any metric in the form (\ref{metric}), the surfaces given by \ $u=$ const. \
are spacelike surfaces on which \ $\d s^2 =-2P^{-2}\d\zeta\d\bar\zeta$. \ 
 The vector fields $P\partial_\zeta$ and $P\partial_{\bar\zeta}$ are tangent
to these surfaces which are orthogonal to the repeated principal null
direction $\ell^\mu$. They are therefore regarded as ``wave surfaces''
associated with waves propagating in the null direction $\ell^\mu$.

One of the field equations immediately implies that 
 \begin{equation}
 P^2(\log P)_{,\zeta\bar\zeta} =\Lambda ,
 \label{Peqn} 
 \end{equation}
 which ensures that the Gaussian curvature of the wave surfaces given by \ 
$K=2P^2(\log P)_{,\zeta\bar\zeta}$ \ is constant with a value that is directly
related to the cosmological constant. Also, it can be shown that a general
solution of (\ref{Peqn}) is given by 
 \begin{equation} 
 P=\big(1+\Lambda F\bar F\big) 
\big(F_\zeta\bar F_{\bar\zeta}\big)^{-1/2} 
 \label{genP}
 \end{equation}  
 where $F=F(\zeta,u)$ is an arbitrary complex function, holomorphic in $\zeta$.
It is then possible to use the coordinate freedom \
$\zeta\to\tilde\zeta(u,\zeta)$ \ to put \ $F=\zeta$ \ so that, in general 
 \begin{equation}
 P=1+\Lambda\zeta\bar\zeta.
 \label{P}
 \end{equation} 
 This form will generally be assumed below.

It may be noted that (\ref{Peqn}) also admits the particular solution 
 \begin{equation}
 P=\sqrt{-\Lambda}\,(\zeta+\bar\zeta), 
 \label{altP}
 \end{equation}
 which is valid only for a negative cosmological constant. However, using tilded
coordinates for (\ref{altP}), this is related to the standard form (\ref{P}) by
the transformation 
 \begin{equation}
\sqrt{-\Lambda}\,\zeta={1-\tilde\zeta\over1+\tilde\zeta}, \qquad\qquad
 \tilde\zeta={1-\sqrt{-\Lambda}\,\zeta\over1+\sqrt{-\Lambda}\,\zeta}.
 \label{zetaPtrans}
 \end{equation}
 The use of the form (\ref{altP}) will be clarified in section~9 below.

\section{Coordinate freedoms}

There are a number of remaining coordinate freedoms that may still be employed to
simplify expressions.

First, there is the ability to alter the origin of the affine parameter on each
geodesic 
 \begin{equation}
 r\to\tilde r=r+R(\zeta,\bar\zeta,u), 
 \label{rtrans}
 \end{equation}
 under which 
 $$ \begin{array}{l}
 {\displaystyle \tilde W\o =W\o -R_{,\zeta}
-{2\bar\tau\o R\over P} }, \\[10pt]
\tilde G\o=G\o+(\tau\o\bar\tau\o+\Lambda)R, \\[6pt]
\tilde H\o=H\o -(\tau\o\bar\tau\o+\Lambda)R^2 -2G\o R -R_{,u}.  
 \end{array} $$

Secondly, there is a freedom in the choice of null parameter 
 \begin{equation}
 u\to\tilde u=w(u), \qquad r\rightarrow\tilde{r}={1\over w_{,u}}r
 \label{utrans}
 \end{equation}
 under which 
 $$ \tilde W={1\over w_{,u}}\,W, \qquad \tilde H={1\over w_{,u}^2}\,H
+{w_{,uu}\over w_{,u}^3}\,r. $$

\goodbreak

Finally, for the choice (\ref{P}) it may be noted that, when $p$ and $q$ are
arbitrary complex functions of $u$, the transformation 
 \begin{equation}
 \zeta\to\tilde\zeta ={\bar q+p\zeta\over\bar p-\Lambda q\zeta}, 
 \label{zetatrans}
 \end{equation}
 leaves the component \ $2P^{-2}\d\zeta\d\bar\zeta$ \ invariant, but induces
changes in $H$ and $W$. Specifically 
 $$ W\o \to \tilde W\o 
={(\bar p-\Lambda q\zeta)^2\over p\bar p+\Lambda q\bar q} \left[ W\o
+{m+in\bar\zeta+\Lambda\bar m\bar\zeta^2\over P^2} \right],  \qquad
\tilde\tau\o =\left( {p-\Lambda\bar q\bar\zeta 
\over\bar p-\Lambda q\zeta} \right)\tau\o, $$ 
 where
 $$ m={(pq_{,u}-qp_{,u})\over p\bar p+\Lambda q\bar q},
\qquad n= i{(\bar pp_{,u}-p\bar p_{,u}+\Lambda\bar qq_{,u}-\Lambda q\bar
q_{,u}) \over p\bar p+\Lambda q\bar q}. $$ 
 Similarly, for the choice (\ref{altP}), we can use the transformation \ 
$\zeta\to\tilde\zeta =p\zeta+iq$, \ where $p$ and $q$ are then real functions
of~$u$.

\section{Integration for $\tau\o$ and its canonical forms}

Let us first consider the Ricci identity \
 $\bar\delta\tau=2\tau(\bar\tau-\bar\beta)+2\Lambda$, \ which now becomes 
 \begin{equation}
 \left(\tau\o\over P\right){}_{,\zeta} ={\tau\o\bar\tau\o+2\Lambda\over P^2}. 
 \label{Rq}
 \end{equation}
 It is immediately clear that $\tau\o$ cannot be zero for a non-zero cosmological
constant. The fact that the right hand side is real implies that there exists a
real function $\phi(\zeta,\bar\zeta,u)$ such that 
 \begin{equation}
 \tau\o=P\,\phi_{,\bar\zeta}. 
 \label{phi}
 \end{equation}
 For the general case in which $P$ is given by (\ref{P}), the Ricci identity \
$\delta\tau=2\tau\beta$ \ can then be integrated to show that 
 \begin{equation}
 \phi=\log P-\log Q,
 \label{phi2}
 \end{equation}
 where \ $Q=a+\bar b\zeta+b\bar\zeta-a\Lambda\zeta\bar\zeta$, \ and $a$ and $b$
are arbitrary real and complex functions of $u$ respectively. We thus obtain  
 \begin{equation}
 \tau\o=  {-b+2a\Lambda\zeta+\bar b\Lambda\zeta^2 \over
a+\bar b\zeta+b\bar\zeta-a\Lambda\zeta\bar\zeta}.
 \label{tauo}
 \end{equation} 
 Let us immediately note that, if \ $b=0$, \ we have 
 \begin{equation}
 {\bf Case \ 1:} \hskip8pc
\tau\o={2\Lambda\zeta\over(1-\Lambda\zeta\bar\zeta)} \hskip12pc
 \label{tauo1}
 \end{equation}
 Similarly, if \ $a=0$, \ we can always make $b$ real and then 
 \begin{equation}
 {\bf Case \ 2:} \hskip8pc
 \tau\o=  -{(1-\Lambda\zeta^2) \over\zeta+\bar\zeta} \hskip12pc
 \label{tauo2}
 \end{equation}
 These clearly reduce to the two standard cases which are well known when
$\Lambda=0$.

It may be seen in general that \
 $\tau\o\bar\tau\o+\Lambda=(b\bar b+\Lambda a^2){P^2\over Q^2}$. \ (This
expression occurs in the metric function $H$.) It is then convenient to
separately label the coefficient 
 \begin{equation}
 k=b\bar b+\Lambda a^2. 
 \label{k}
 \end{equation}
 This has been identified by Ozsv\'ath, Robinson and R\'ozga \cite{OzRoRo85} as
a quantity of invariant sign which can be used to assist in the classification of
this family of solutions.

The expression (\ref{tauo}) can now be simplified by use of the transformation
(\ref{zetatrans}). This leaves its general form invariant, but with coefficients
transformed as 
 $$ a\to \tilde a = (p\bar p-\Lambda q\bar q)a 
+\bar pq\,b +p\bar q\,\bar b, \qquad\qquad
 b\to \tilde b = -2\Lambda\bar p\bar q\,a +\bar p^2b 
-\Lambda\bar q^2\bar b  $$ 
 so that \ $k\to\tilde k=(p\bar p+\Lambda q\bar q)^2k$. \  The character of this
transformation, however, depends on the invariant signs of the quantities
$\Lambda$ and $k$.

\medskip\goodbreak\noindent
{\bf When $\Lambda=0$}, \ $k=b\bar b$. \ 
In this case, if \ $b\ne0$, \ it is always possible to use the above
transformation to put \ $a=0$, \ so that (\ref{tauo}) can be transformed to
(\ref{tauo2}) and we arrive at case 2 which, for type~N, are the Kundt waves. \ If
\ $b=0$, \ then \ $k=0$ \ and \ $\tau\o=0$ \ and, for type~N, these are the {\sl
pp}-waves. \ This identifies two canonical types for these solutions: 
 $$ \Lambda=0 \qquad \cases{ 
k=0 \qquad \mbox{generalised {\sl pp}-waves \hskip 3.9pc case 1}  \cr
\noalign{\medskip}
k>0 \qquad \mbox{generalised Kundt waves \hskip 2.3pc case 2}  } $$ 
 These are distinct solutions, which cannot be related to each other.

\medskip\goodbreak\noindent
{\bf When $\Lambda>0$}, \ it is only possible for $k$ to be positive. \ 
In this case, it is always possible to choose the function $p\over q$ to
put either \ $\tilde a=0$ \ and \ $\tilde b\ne0$, \ or \ $\tilde a\ne0$ \
and \ $\tilde b=0$. \ A further transformation using the freedom in $p$,
with $q=0$, can then be used to set either $b=1$ or $a=1$ respectively. \ 
Thus, $\tau\o$ can always be transformed into either of the canonical forms
(\ref{tauo1}) or (\ref{tauo2}). Indeed, it is possible to transform (\ref{tauo1})
into (\ref{tauo2}) or vice versa. These two forms, for this case, are
completely equivalent. This family of solutions may therefore be considered
to be a generalisation of either the Kundt waves or the {\sl pp}-waves in
the sense that they reduce to either of these forms for type~N solutions in
the appropriate limit depending on the particular coordinates adopted. 
 $$ \Lambda>0 \quad \Rightarrow \quad k>0 \qquad
 \mbox{generalised {\sl pp} and Kundt waves 
\hskip 2.5pc cases 1 \& 2}   $$

\medskip\goodbreak\noindent
{\bf When $\Lambda<0$}: 
In this case there are three distinct possibilities which are identified
by the sign of~$k$. 

\noindent
1. If \ $k>0$, \ then \ $b\bar b>-\Lambda a^2$, \ and it is always possible
to use the above transformation to put \ $a=0$ \ and hence to obtain
generalised Kundt waves (case 2). 

\noindent
2. Similarly, if $k<0$, \ $-\Lambda a^2>b\bar b$ \ and it is always possible
to use the transformation to put \ $b=0$ \ to obtain generalised {\sl
pp}-waves (case 1). 

\noindent
3. However, another distinct case arises here when \ $k=0$. \ This occurs,
when \ $b=\sqrt{-\Lambda}e^{i\theta}a$ \ for any arbitrary
function $\theta(u)$. In this case, the above transformation reduces to
the following form:
 $$ \tilde a=(p+\sqrt{-\Lambda}e^{i\theta}q)
(\bar p+\sqrt{-\Lambda}e^{-i\theta}\bar q)\,a \qquad\qquad
\tilde b=\sqrt{-\Lambda}e^{i\theta} 
(\bar p+\sqrt{-\Lambda}e^{-i\theta}\bar q)^2a  $$ 
 so that it is obviously not possible to reduce either $a$ or $b$
separately to zero. This particular case corresponds to that in which $Q$
factorises, and hence $\tau\o$ is given by 
 $$ \tau\o=  -\sqrt{-\Lambda}e^{i\theta} \left(
{1+\sqrt{-\Lambda}e^{-i\theta}\zeta \over 
1+\sqrt{-\Lambda}e^{i\theta}\bar\zeta} \right). $$ 
 However, it is clearly possible to remove the phase $e^{i\theta}$, so that this
additional case can be expressed in the canonical form 
 \begin{equation} 
 {\bf Case \ 3}_{(\Lambda<0)}: \hskip4pc
 \tau\o=  -\sqrt{-\Lambda} \left( {1+\sqrt{-\Lambda}\,\zeta \over 
1+\sqrt{-\Lambda}\,\bar\zeta} \right). \hskip8pc 
 \label{tauo3}
 \end{equation} 
 These solutions generalise those of Siklos \cite{Siklos85} using a different
coordinate system. 

\noindent
Thus, in this case, there exist three distinct canonical types: 
 $$ \Lambda<0 \qquad \cases{ 
k<0 \qquad \mbox{generalised {\sl pp}-waves \hskip 4.4pc case 1}  \cr
\noalign{\medskip}
k=0 \qquad \mbox{generalised Siklos waves \hskip 3pc case 3}  \cr
\noalign{\medskip}
k>0 \qquad \mbox{generalised Kundt waves \hskip 2.7pc case 2}  } $$

\medskip\goodbreak
For an analysis of local exact solutions, it is sufficient to consider separately
the three particular cases listed above: i.e. $P$ is given by (\ref{P}) and
$\tau$ by (\ref{tauo1}), (\ref{tauo2}) or (\ref{tauo3}). However, since $a(u)$ and
$b(u)$ in (\ref{tauo}) are arbitrary functions, it is possible also to construct
composite space-times in which these functions are non-zero for different ranges
of $u$. In the remainder of this paper, however, equations are expressed in forms
which apply to any of the particular cases above. It is only assumed that
$\tau\o$ is independent of $u$.

\section{Integration for $G\o$ and $W\o$}

The metric equations imply that $\gamma$ and $\mu$ have the forms \
$\gamma=-(\tau\o\bar\tau\o+\Lambda)r+G\o+{1\over2}im\o$ \ and \ $\mu=im\o$ \
where 
 \begin{equation}
  m\o= {\textstyle{1\over2}} iP^2\big[ W\o_{\ ,\bar\zeta}
-\bar W\o_{\ ,\zeta} \big]
+iP(\tau\o W\o-\bar\tau\o\bar W\o).
 \end{equation} 
 The $r$ component of the Ricci identity \ $\delta\gamma-\Delta\beta=\mu\tau$ \
for this case vanishes identically, and the remaining part becomes 
 $$ \left( G\o +{\textstyle{1\over2}}im\o \right)_{,\bar\zeta}
=-(\tau\o\bar\tau\o+\Lambda)\bar W\o +{im\o\tau\o\over P}. $$ 
 Using the Ricci identity \ $(P\tau\o)_{,\bar\zeta}={\tau\o}^2$, \ this equation
can be rewritten in the form 
 \begin{equation}
 \left[G\o 
-{P^2\over4} \left[ W\o_{\ ,\bar\zeta} -\bar W\o_{\ ,\zeta} \right] 
+{P\over2}(\tau\o W\o+\bar\tau\o\bar W\o) \right]_{,\bar\zeta} 
 ={\tau\o P\over2}\left[
W\o_{\ ,\bar\zeta} +\bar W\o_{\ ,\zeta} \right] -\Lambda\bar W\o.
 \label{Ro}
 \end{equation}
 At this point, we can use the coordinate freedom (\ref{rtrans}) to set 
 \begin{equation}
 W\o_{\ ,\bar\zeta} +\bar W\o_{\ ,\zeta} =0.
 \label{omegaconstraint}
 \end{equation}
 It can be seen that any further freedom in $R$ must be constrained by the
condition  
 \begin{equation}
 P^2R_{,\zeta\bar\zeta} 
+P(\tau\o R_{,\zeta}+\bar\tau\o R_{,\bar\zeta})
+2(\tau\o\bar\tau\o+2\Lambda)R=0.  
 \label{Rcond}
 \end{equation} 
 The condition (\ref{omegaconstraint}) implies that there exists a real
function $X(\zeta,\bar\zeta,u)$ such that 
 $$ W\o=-i\,X_{,\zeta}. $$
 With this, the identity (\ref{Ro}) can be integrated to give 
 \begin{equation}
 G\o +{\textstyle{1\over2}}P(\tau\o W\o+\bar\tau\o\bar W\o)
 +{\textstyle{1\over2}}i\left[P^2X_{,\zeta\bar\zeta}+2\Lambda X\right]
 =-2\Lambda f(\zeta,u), 
 \label{Roint}
 \end{equation} 
 where $f(\zeta,u)$ is an arbitrary function, and the factor $-2\Lambda$ has
been added for convenience. (It may be noted that any function $f$ which is
independent of $\zeta$ may be removed by an appropriate transformation.) The real
and imaginary parts of (\ref{Roint}) are respectively 
 \begin{equation}
 G\o =-{P\over2}(\tau\o W\o+\bar\tau\o\bar W\o) -\Lambda(f+\bar f) 
 \label{Go}
 \end{equation} 
 and
 $$ P^2X_{,\zeta\bar\zeta}+2\Lambda X =2i\Lambda(f-\bar f), $$ 
 but the remaining freedom (\ref{rtrans}) satisfying (\ref{Rcond}) can be used
(provided
$\tau\o\ne0$) to set \ $X_{,\zeta\bar\zeta}=0$. \ Thus, provided $\Lambda\ne0$,
we can set \ $X=i(f-\bar f)$ \ and 
 \begin{equation}
 W\o=f_{,\zeta}, 
 \label{Wo}
 \end{equation} 
 which is independent of $\bar\zeta$. In fact, except for the case~1 solutions
with $\Lambda=0$, these solutions are described by the arbitrary holomorphic
function $W\o(\zeta,u)$ (or $f(\zeta,u)$), together with the arbitrary real
function $H\o(\zeta,\bar\zeta,u)$.

\section{The curvature tensor components}

With the above conditions, we now have \ 
$\mu=im\o=-P(\tau\o W\o-\bar\tau\o\bar W\o)$. \ The only spin coefficient which
has not yet been specified is $\nu$, which can be expressed as \
$\nu=n\o r+\nu\o$, \ where 
 $$ n\o = -\tau\o(P^2W\o)_{,\zeta} -2\Lambda PW\o 
+P\bar\tau\o(\tau\o W\o-\bar\tau\o\bar W\o) $$
 and
 $$ \nu\o = PH\o_{,\zeta} +2\bar\tau\o H\o -PW\o_{,u} 
 +P^2W\o (\tau\o W\o+\bar\tau\o\bar W\o) +2\Lambda PW\o(f+\bar f). $$ 
 The remaining Ricci identities give expressions for the non-zero components of
the curvature tensor. After substituting the above relations, we obtain 
 \begin{eqnarray}
&& \begin{array}{l}
  \Psi_3 = -\tau\o(P^2W\o)_{,\zeta} -2\Lambda PW\o  
 \end{array}
\label{Psi3}\\[6pt]
&& \begin{array}{l}
  \Psi_4 = \Big[ -P\tau\o(P^2W\o)_{,\zeta\zeta}
 +2(\tau\o\bar\tau\o-\tau\o P_{,\zeta}-2\Lambda)(P^2W\o)_{,\zeta}  
 +6\Lambda P\bar\tau\o W\o  \Big]r \\[6pt]
 \qquad\qquad +P^2H\o_{,\zeta\zeta} +2P(\bar\tau\o+P_{,\zeta})H\o_{,\zeta}
+2\bar\tau\o{}^2H\o -(P^2W\o)_{,u\zeta} \\[6pt] 
 \qquad\qquad +\big[ 3P\tau\o W\o+P\bar\tau\o\bar W\o
+2\Lambda(f+\bar f)\big] (P^2W\o)_{,\zeta} \\[6pt]
 \qquad\qquad +2P^2\bar\tau\o{}^2W\o\bar W\o +6\Lambda P^2W\o{}^2  
 \end{array}\\[6pt]
&& \begin{array}{l}
  \Phi_{22} = P^2H\o_{,\zeta\bar\zeta} 
+P\tau\o H\o_{,\zeta} +P\bar\tau\o H\o_{,\bar\zeta}
+2(\tau\o\bar\tau\o+2\Lambda)H\o \\[6pt] 
 \qquad\qquad +P\tau\o\bar W\o(P^2W\o)_{,\zeta}
+P\bar\tau\o W\o(P^2\bar W\o)_{,\bar\zeta} 
+2P^2(\tau\o\bar\tau\o+3\Lambda)W\o\bar W\o
 \end{array}
 \end{eqnarray}
 In these expressions, $P$ is given by (\ref{P}), and $\tau\o$ can take any of
its canonical forms (\ref{tauo1}), (\ref{tauo2}) or (\ref{tauo3}). They contain
an arbitrary holomorphic function $W\o(\zeta,u)$ (or $f(\zeta,u)$), and an
arbitrary real function $H\o(\zeta,\bar\zeta,u)$. For case~2, these reduce to
the known expressions for the Kundt waves in the limit as $\Lambda\to0$.

For the case~1 solutions, the limit as $\Lambda\to0$ of the above expressions only
leads to the type~N {\sl pp}-waves. Since $\tau\o=0$ in this limit, it is not
possible as above to set $W\o$ independent of $\bar\zeta$. In this case, however,
the transformation (\ref{rtrans}) can alternatively be used to set \
$W\o=W\o(\bar\zeta,u)$ \ and the curvature tensor components with $\Lambda=0$
and $\tau\o=0$ can be expressed as  
 \begin{eqnarray}
 \begin{array}{l}
\Psi_3 ={\textstyle{1\over2}}\,\bar W\o_{\ ,\zeta\zeta} \\[6pt] 
 \Psi_4 ={\textstyle{1\over2}}\,\bar W\o_{\ ,\zeta\zeta\zeta}\,r 
+H\o_{\ ,\zeta\zeta} -W\o\bar W\o_{\ ,\zeta\zeta} \\[6pt] 
 \Phi_{22} =H\o_{\ ,\zeta\bar\zeta}
-{\textstyle{1\over2}}(W\o W\o_{\ ,\bar\zeta\bar\zeta}
+\bar W\o\bar W\o_{\ ,\zeta\zeta})
-{\textstyle{1\over2}}(W\o_{\ ,u\bar\zeta}+\bar W\o_{\ ,u\zeta})
-{\textstyle{1\over2}}\left ({W\o_{\ ,\bar\zeta}}^2 
+\bar W\o_{\ ,\zeta}{}^2 \right) 
 \end{array}
 \end{eqnarray}

\section{The type N reductions with $\Lambda\ne0$}

It may first be noticed that the expression (\ref{Psi3}) for $\Psi_3$ can be
written as 
 \begin{equation}
 \Psi_3 = -{P^2\over Q}\big( Q\tau\o W\o \big){}_{,\zeta} .
 \end{equation}
  This clearly vanishes (and the solution reduces to type~N) when \ $Q\tau\o
W\o=c\Lambda$, \ where $c$ is an arbitrary complex function of $u$, so that 
 \begin{equation}
 W\o={c\Lambda\over -b+2a\Lambda\zeta+\bar b\Lambda\zeta^2}. 
 \label{WoN}
 \end{equation}  
 The multiple $\Lambda$ has been included since this term can obviously be
transformed away when $\Lambda=0$. Further, when $c$ is real and $k\ne0$, the
transformation (\ref{rtrans}) can be used to set both $W\o$ and $G\o$ to zero
simultaneously. Thus, for case~1 and case~2 solutions, $c$ may be purely
imaginary. However, $c$ may be generally complex for the case~3 solutions which
occur when $\Lambda_c<0$.

For this type~N limit, it can be shown that the $r$ component of $\Psi_4$
vanishes, as is required by the Bianchi identities, and general expressions for
the non-zero components of the curvature tensor are 
 \begin{eqnarray}
&& \begin{array}{l}
  \Psi_4 = P^2H\o_{,\zeta\zeta} +2P(\bar\tau\o+P_{,\zeta})H\o_{,\zeta}
+2\bar\tau\o{}^2H\o -(P^2W\o)_{,u\zeta} \\[6pt] 
 \qquad\qquad {\displaystyle
+2P^2{\bar\tau\o\over\tau\o}(\tau\o\bar\tau\o-\Lambda)W\o\bar W\o
 +2\Lambda(f+\bar f)(P^2W\o)_{,\zeta}  } 
 \end{array}\\[6pt]
&& \begin{array}{l}
  \Phi_{22} = P^2H\o_{,\zeta\bar\zeta} 
+P\tau\o H\o_{,\zeta} +P\bar\tau\o H\o_{,\bar\zeta}
+2(\tau\o\bar\tau\o+2\Lambda)H\o 
+{\displaystyle{2kP^4\over Q^2}W\o\bar W\o}  
 \end{array}
 \end{eqnarray} 
 with $W\o$ given by (\ref{WoN}).  When $W\o=0$ (i.e. when $c=0$) and
$\Lambda\ne0$, these are the solutions that are described in \cite{OzRoRo85} and
\cite{BicPod99a}. However, when $c\Lambda\ne0$, this identifies a new family of
type~N solutions that has not been identified in previous literature.

\section{Alternative form for $P$ when $\Lambda<0$}

Let us now consider the special case in which $\Lambda<0$ and $P$ is given by
(\ref{altP}), i.e. $P=\sqrt{-\Lambda}\,(\zeta+\bar\zeta)$. \ In
this case, the integration for $\tau\o$ can be performed in a similar way to
that above, and we obtain 
 \begin{equation} 
 \tau\o ={\sqrt{-\Lambda}
(\tilde a+2i\tilde b\zeta-\tilde c\zeta^2)\over
\tilde a+i\tilde b\zeta-i\tilde b\bar\zeta+\tilde c\zeta\bar\zeta}, 
 \label{tau2}
 \end{equation}  
 where $\tilde a$, $\tilde b$ and $\tilde c$ are arbitrary real functions of $u$.
We can then use the coordinate freedom $\zeta\to\tilde\zeta =p\zeta+iq$, \ where
$p$ and $q$ are real functions of $u$ to simplify this expression. Specifically,
if $\tilde c\ne0$, we can set 
 \begin{equation}
 {\bf Case \ 4:} \hskip8pc 
\tau\o ={\sqrt{-\Lambda}(\epsilon-\zeta^2) 
\over\epsilon+\zeta\bar\zeta}, \hskip9pc 
 \label{tauo4}
 \end{equation} 
 where $\epsilon=1,0,-1$. Alternatively, if $\tilde c=0$ and $\tilde b\ne0$, we
can set 
 \begin{equation}
 {\bf Case \ 5:} \hskip8pc \tau\o
={2\sqrt{-\Lambda}\,\zeta\over\zeta-\bar\zeta}. \hskip11pc 
 \label{tauo5}
 \end{equation} 
 Finally, if $\tilde c$ and $\tilde b$ both vanish, we obtain 
 \begin{equation}
 {\bf Case \ 6:} \hskip8pc \tau\o=\sqrt{-\Lambda}. \hskip12pc 
 \label{tauo6}
 \end{equation}

These forms, however, can be related directly to those above using the
transformation (\ref{zetaPtrans}). This yields the expression (\ref{tauo}) for
$\tau\o$ where 
 $$ a=\tilde a+\tilde c, \qquad 
b=\sqrt{-\Lambda}(\tilde a-\tilde c+2i\tilde b). $$

\smallskip\noindent
For {\bf case 4}$_{(\Lambda<0)}$, we have \ $\tilde a=\epsilon$, \ $\tilde
b=0$, \ $\tilde c=1$. \hfill\break It is therefore evident that \
$a=\epsilon+1$ \ and \ $b=\sqrt{-\Lambda}(\epsilon-1)$. \ Thus \hfil\break
 \indent -- \qquad the case when \ $\epsilon=1$ \ belongs to {\bf case 1}
with $\Lambda<0$. \hfil\break
 \indent -- \qquad the case when \ $\epsilon=0$ \ corresponds to {\bf case
3}$_{(\Lambda<0)}$. \hfil\break
 \indent -- \qquad the case when \ $\epsilon=-1$ \ belongs to {\bf case 2}
with $\Lambda<0$.

\smallskip\noindent
For {\bf case 5}$_{(\Lambda<0)}$, we have \ $\tilde a=0$, \ $\tilde
b=1$, \ $\tilde c=0$, so that \ $a=0$ \ and \ $b=2i\sqrt{-\Lambda}\>$.
\hfil\break
 \indent -- \qquad This case belongs to {\bf case 2} with $\Lambda<0$, but in a
form in which $\zeta$ is replaced by $i\zeta$.

\smallskip\noindent
For {\bf case 6}$_{(\Lambda<0)}$, we have \ $\tilde b=0$, \ and \ $\tilde
c=0$, so that \ $a=\tilde a$ \ and \
$b=\sqrt{-\Lambda}\,\tilde a\>$. \hfil\break
 \indent -- \qquad This case corresponds to {\bf case 3}$_{(\Lambda<0)}$.

\smallskip
It may be noted that the apparent difference between
case~4 with $\epsilon=0$ and case~6 corresponds only to the phase of $\zeta$
which may always be reduced to zero. Similarly, a transformation exists which
relates case~4 with $\epsilon=-1$ and case~5. (The equivalence of cases
4$_{(\epsilon=-1)}$ and 5 is similar to the equivalence of cases 1 and 2 when
$\Lambda>0$.) Thus it is sufficient to consider just three distinct canonical
types. These can be taken as the three types of case~4 with different values of
$\epsilon$ or, preferably, case~4 with $\epsilon=1$, case~5 and case~6. These
canonical types were identified by Siklos \cite{Siklos85}, although he only
investigated case~6.

\section{Interpretation of the solutions} 

The solutions described above are radiative space-times in which the rays are
non-expanding (as well has having zero shear and twist). Moreover, the wave
surfaces have constant curvature proportional to the cosmological constant.
However, the fact that $\tau$ is non-zero indicates that subsequent wave surfaces
are locally rotated relative to each other. These general properties will be
illustrated below.

Apart from the quantities $\Lambda$, $P$ and $\tau\o$, the above family of
solutions is essentially represented by an arbitrary holomorphic function
$W\o(\zeta,u)$ (or $f(\zeta,u)$), and an arbitrary real function
$H\o(\zeta,\bar\zeta,u)$. Our general method of approach will be to initially
consider the weak field limit in which $H\o$ and $W\o$ (and $f$) are taken to be
zero. This identifies the background in which weak fields of this type propagate.
In this limit, the space-time is conformally flat and is Minkowski,
de~Sitter or anti-de~Sitter according as the cosmological constant is
respectively zero, positive or negative.

The background metric is given by (\ref{metric}) with 
 \begin{equation}
 H=-(\tau\o\bar\tau\o+\Lambda)r^2, \qquad W={2\bar\tau\o\over P}\,r 
 \label{cflimit}
 \end{equation} 
 for different values of $\Lambda$, and hence differing expressions for $P$ and
$\tau\o$. In these background space-times, it is possible to explicitly
investigate the geometry of the waves surfaces and the way in which they foliate
the space-time. This will also enable us to investigate the location and
character of the singularities of the solutions in this weak field background.

This approach is also appropriate in the study of exact sandwich waves as it
explicitly determines the geometries of the shock fronts and backs of the waves.

\section{Waves in a Minkowski background}

We first consider the case in which $\Lambda=0$. Here, space-times of case~1, for
which $\tau=0$, are the {\sl pp}-waves which have plane wave surfaces and parallel
rays. They include plane waves. These solutions are well known and need not be
further discussed here.

We therefore only need consider case~2, for which \ 
$\tau\o=-(\zeta+\bar\zeta)^{-1}$. \ Putting \ $\zeta={1\over\sqrt2}(x+iy)$,
\ the line element (\ref{metric}) in the conformally flat limit (\ref{cflimit})
with $\Lambda=0$ takes the form 
 \begin{equation}
 \d s^2=2\d u\left(\d r-{r^2\over2x^2}\,\d u
-{2\>r\over x}\,\d x\right) -\d x^2 -\d y^2, 
 \label{Minkowski}
 \end{equation} 
 which is related to the cartesian form of Minkowski space 
 $$ \d s^2=\d T^2-\d X^2-\d Y^2-\d Z^2 $$ 
 by the transformation 
 \begin{equation}
 \begin{array}{l}
{\displaystyle T={1\over2x}\,\left(r+u^2r+2ux^2\right)} \\[12pt] 
{\displaystyle Z={1\over2x}\,\left(r-u^2r-2ux^2\right)} \\ [12pt]
{\displaystyle X=x+{u\,r\over x}} \\[12pt]
 Y=y 
 \end{array} \qquad \qquad
 \begin{array}{l}
{\displaystyle u= {X\mp\sqrt{X^2+Z^2-T^2}\over T+Z}} \\[12pt]
{ r=\pm(T+Z)\sqrt{X^2+Z^2-T^2}} \\[12pt]
{\displaystyle x= \pm\sqrt{X^2+Z^2-T^2}} \\[12pt]
 y=Y 
 \end{array}
 \label{Mtrans}
 \end{equation}

It can be seen from (\ref{Mtrans}) that the wave surfaces \ $u=u_0=$~const. \ in
this background space-time are given by 
 $$ (1+u_0^2)T-2u_0X-(1-u_0^2)Z=0, $$ 
 which describes a family of null hyperplanes whose orientation varies for
different values of $u_0$. Putting \ $u_0=\tan(\alpha/2)$, \ where $\alpha$
is a constant on any wave surface, these can be written in the form
 \begin{equation}
 \sin\alpha\,X +\cos\alpha\,Z =T. 
 \label{plane}
 \end{equation}
 This clearly demonstrates that, at any time, successive wave surfaces 
$u=u_0$ are rotated about the $Y$ axis as $u_0$ increases from $-\infty$ to
$+\infty$ (or as $\alpha$ goes from $-\pi$ to $+\pi$). The rotation of these
planes for different values of $u_0$ is consistent with the non-zero value of
$\tau$ for these metrics.

For all this family of solutions, it may be noticed that the curvature components
$\Psi_3$, $\Psi_4$ and $\Phi_{22}$ are unbounded when $\tau\o$ is unbounded. This
occurs when \ $\zeta+\bar\zeta=0$ \ (or $x=0$ in the above notation). In the
background space-time, this singularity occurs on the hypersurface 
 \begin{equation}
 X^2+Z^2=T^2, 
 \label{cylinder}
 \end{equation}
 which is an expanding cylinder centred on the $Y$-axis and whose radius expands
with the speed of light. It may be noticed that each wave surface touches this
cylinder on the line \ $X=\sin\alpha\,T$, \ $Z=\cos\alpha\,T$. \ Thus, the
singularity on the expanding cylinder (\ref{cylinder}) can be interpreted as the
caustic formed from the envelope of the family of wave surfaces.

It may also be observed that two plane wave surfaces (\ref{plane}) pass through
each point outside the cylinder (\ref{cylinder}). However, this repetition must be
excluded according to the assumption that the tangent vector $\ell^\mu$ is unique.
The complete family of wave surfaces should therefore be taken as the family of
half-planes for which $x\ge0$. These are illustrated in figure~1. The geometry of
these wave surfaces in the background space-time is exactly the same as that of
the pure radiation conformally flat case described in \cite{GriPod98}.

\begin{figure}[hpt]
\begin{center} \includegraphics[scale=0.4, trim=5 5 5 -5]
{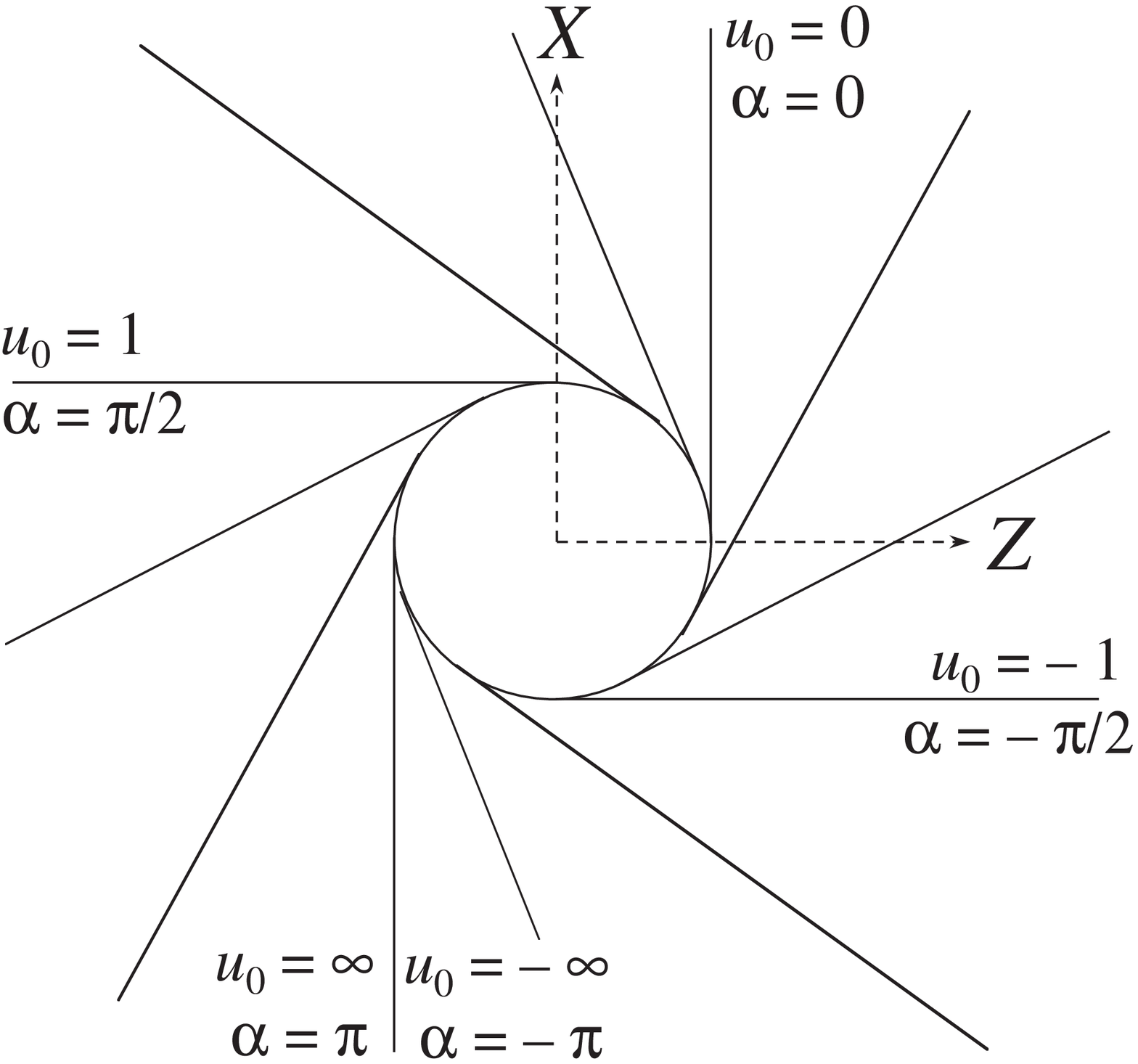} \qquad
\includegraphics[scale=1.0, trim=5 -5 5 -5]
{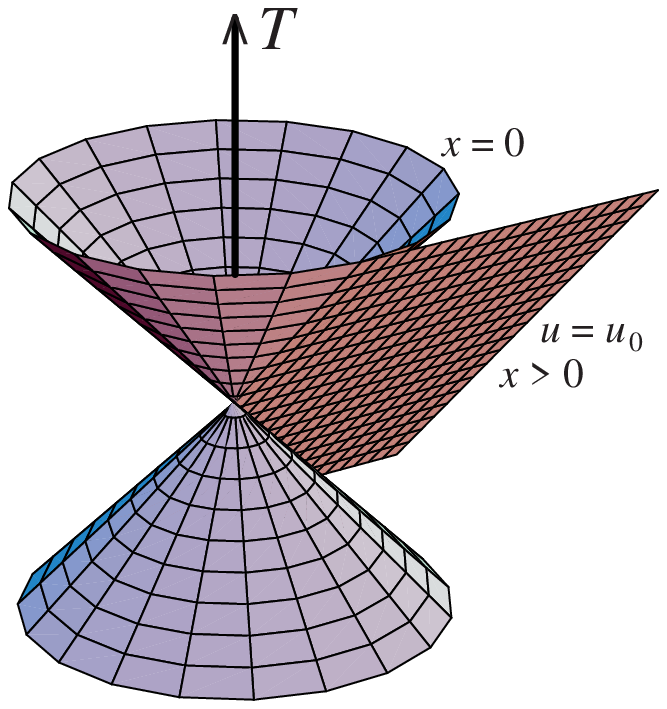}
\caption{ \small The geometry of the wave surfaces in the background Minkowski
space-time. \hfil\break
(Left) For constant $T$ and $Y$, the wave surfaces $u=u_0=$~const. can be
represented as a family of half-lines at a perpendicular distance $T$ from an
origin as indicated. The envelope of the lines forms a circle corresponding to the
coordinate singularity at $x=0$. As $T$ increases, the circle expands and each
wave surface propagates perpendicular to the tangent to that surface. \hfil\break
(Right) The half-plane wave surfaces are tangent to a 2-dimensional null cone on
which $x=0$. The picture illustrates the singularity and a single wave surface in
a section with constant $Y$.  }
\end{center}
\end{figure}

It may be observed that no wave surfaces pass through points that are inside the
expanding cylinder (\ref{cylinder}). The coordinates used in (\ref{metric}) do
not cover this part of the space-time.

Any observer outside the cylindrical envelope could detect a gravitational wave
whose direction of propagation will rotate until it is reached by the singularity
itself.

For this family of solutions, the scalar polynomial invariants all vanish. The
singularity which occurs when $\zeta+\bar\zeta=0$ is therefore not a polynomial
curvature singularity. Nevertheless, some curvature tensor components diverge and
an observer approaching this singularity would experience unbounded tidal forces.

Finally, it should be pointed out that the complete solution is time-symmetric,
so that the envelope of wave surfaces is a cylinder whose radius decreases to zero
at the speed of light and then increases.

\section{The de~Sitter and anti-de~Sitter backgrounds for case 2 solutions}

Let us next consider the case 2 solutions in which $\Lambda\ne0$ and $\tau\o$ has
the canonical form (\ref{tauo2}). In this case, the line element (\ref{metric})
in the conformally flat limit (\ref{cflimit}) takes the form  
 $$ \d s^2 =2\d u\left[ \d r
-{(1+\Lambda\zeta\bar\zeta)^2r^2\over(\zeta+\bar\zeta)^2}\d u
-{2(1-\Lambda\bar\zeta^2)r\over (1+\Lambda\zeta\bar\zeta)(\zeta+\bar\zeta)}\d\zeta
-{2(1-\Lambda\zeta^2)r\over
(1+\Lambda\zeta\bar\zeta)(\zeta+\bar\zeta)}\d\bar\zeta \right] 
-{2\,\d\zeta\,\d\bar\zeta\over(1+\Lambda\zeta\bar\zeta)^2}. $$ 
 It is then convenient to put \ $r=(Q^2/P^2)v$ \ (with $Q=\zeta+\bar\zeta$)
and \ $\zeta={1\over\sqrt2}(x+iy)$, \ so that the background line element (which
is de~Sitter or anti-de~Sitter space) can be expressed as 
 \begin{equation}
 \d s^2 ={4x^2\over P^2}\,\big(\d u\,\d v -v^2\,\d u^2\big)
-{1\over P^2}\,(\d x^2+\d y^2). 
 \label{case2background}
 \end{equation}

Now, it is well known that the (anti-)de~Sitter space can be represented as
a four-dimensional hyperboloid 
 \begin{equation}
 {Z_0}^2 -{Z_1}^2 -{Z_2}^2 -{Z_3}^2 -\varepsilon{Z_4}^2 =-\varepsilon a^2, 
 \label{hyp}
 \end{equation}
 embedded in a five-dimensional Minkowski space 
 \begin{equation} 
 \d s^2= \d{Z_0}^2 -\d{Z_1}^2 -\d{Z_2}^2 -\d{Z_3}^2 
-\varepsilon\d{Z_4}^2, 
 \label{5dMink}
 \end{equation} 
 where \ $a^2=1/(2|\Lambda|)$, \ $\varepsilon=1$ for a de~Sitter background
($\Lambda>0$), and $\varepsilon=-1$ for an anti-de~Sitter background
($\Lambda<0$). \ The two forms of the metric (\ref{case2background}) and
(\ref{5dMink}) are related by the transformation 
 \begin{equation}
 \begin{array}{l}
{\displaystyle Z_0= {x\over\sqrt2\,P}(v+2u+2u^2v)} \\[12pt] 
{\displaystyle Z_1= {x\over\sqrt2\,P}(v-2u-2u^2v)} \\ [12pt]
{\displaystyle Z_2= {x\over P}(1+2uv)} \\[12pt]
{\displaystyle Z_3= {y\over P}} \\[12pt]
{\displaystyle Z_4= a{(2-P)\over P}} 
 \end{array} \qquad \Leftrightarrow \qquad
 \begin{array}{l}
{\displaystyle u= \pm{1\over\sqrt2}\>
{Z_2-\sqrt{-{Z_0}^2+{Z_1}^2+{Z_2}^2}\over Z_0+Z_1}} \\[12pt]
{\displaystyle v= \pm{1\over\sqrt2}\>
{Z_0+Z_1\over\sqrt{-{Z_0}^2+{Z_1}^2+{Z_2}^2}}} \\[18pt]
{\displaystyle x= \pm{2a\sqrt{\varepsilon a^2-{Z_3}^2
-\varepsilon {Z_4}^2}\over a+Z_4}} \\[12pt] 
{\displaystyle y= {2a\,Z_3\over a+Z_4}} 
 \end{array} 
 \end{equation}
 where \ $P=1+{1\over4a^2}(x^2+y^2)$.

It can immediately be seen that the singularity which occurs when
$\zeta+\bar\zeta=0$ (or~$x=0$) is located on  
 $$ {Z_3}^2+\varepsilon{Z_4}^2=\varepsilon a^2 \qquad {\rm and} \qquad 
{Z_1}^2+{Z_2}^2={Z_0}^2. $$ 
 For the de~Sitter background (for which $\varepsilon=1$), this is an expanding
torus. The sections in the $Z_1,Z_2$ plane are circles which are expanding at the
speed of light with the time coordinate $Z_0$, and sections in the $Z_3,Z_4$
plane are circles of constant circumference $2\pi a$ that may be considered to
be the circumference of the universe. (This is exactly as expected since the
expanding cylinder described above in Minkowski space must become an expanding
torus in the closed de~Sitter space.)

For the anti-de~Sitter background ($\varepsilon=-1$), however, the singularity is
located on an expanding hyperboloid in which the sections in the $Z_1,Z_2$ plane
are circles which are expanding at the speed of light with the time coordinate
$Z_0$, and sections in the $Z_3,Z_4$ plane are hyperbolae. (Again, this is the
negative curvature equivalent of the expanding cylinder.)

For both cases, it can also be seen that the wave surfaces $u=u_0$ are given in
this five-dimensional representation by the intersection of the hyperboloid
(\ref{hyp}) with the plane 
 $$ (1+2u_0^2)Z_0 -(1-2u_0^2)Z_1 \mp2\sqrt2\,u_0\,Z_2 =0. $$ 
 Putting \ $\sin\alpha=\pm{2\sqrt2\,u_0\over1+2u_0^2}$, \ 
$\cos\alpha={1-2u_0^2\over1+2u_0^2}$, \ this becomes 
 $$ \cos\alpha\,Z_1+\sin\alpha\,Z_2 =Z_0. $$ 
 These are a family of planes which rotate relative to each other in the
$Z_1,Z_2$ plane. They cut the four-dimensional hyperboloid at 
 $$ (\sin\alpha\,Z_1-\cos\alpha\,Z_2)^2 +{Z_3}^2 +\varepsilon{Z_4}^2
=\varepsilon a^2. $$

\begin{figure}[hpt]
\begin{center} \includegraphics[scale=0.8, trim=5 5 5 -5]
{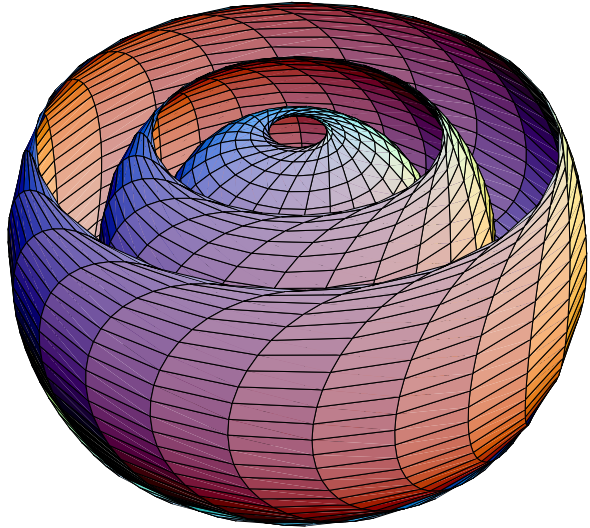} 
\caption{ \small Portions of the de~Sitter universe covered by the coordinates
of (\ref{case2background}) for three values of the time coordinate $Z_0$. Using
$Z_1,Z_2,Z_3$ coordinates ($Z_3$ being the symmetry axis) with $Z_4=0$, the
expanding 3-sphere of the universe reduces to an expanding 2-sphere. The wave
surfaces $u=u_0$ are represented here by semicircles of constant radius~$a$ which
are all tangent to the two expanding circles at $Z_3=\pm a$ which are particular
sections of the expanding torus corresponding to the singularity $x=0$. It can be
seen that the wave surfaces cover a decreasing portion of each successive
2-sphere. }
\end{center}
\end{figure}

For the de~Sitter background ($\varepsilon=1$), these intersections representing
the wave surfaces are a family of spheres with constant area~$4\pi a^2$. Moreover,
the plane cuts are clearly tangent to the expanding torus, so that the singularity
can again be interpreted as a caustic formed from the envelope of wave surfaces.
Also, since two spheres pass through each point within the region covered by these
coordinates, it is appropriate to restrict the family of wave surfaces to the
hemispheres on which $x\ge0$ whose boundary is located on the expanding torus.
(The situation is obviously the positive curvature equivalent of the half-plane
wave surfaces for the Kundt waves in a Minkowski background.) This is illustrated
in figure~2. (Notice that the view along the axis corresponds to the first picture
in figure~1.)

For the anti-de~Sitter background ($\varepsilon=-1$), the wave surfaces
are hyperboloidal. They are tangent to the singularity which is an
expanding hyperboloid that can again be interpreted as an envelope of wave
surfaces. Also, since it is only possible for one wave surface to pass
through any point, it is appropriate to take the wave surfaces as the
family of semi-infinite hyperboloids on which $x\ge0$ (obviously generalising
the half-planes of the Minkowski background).

For this case ($\Lambda<0$), an additional apparent singularity occurs when $P=0$,
i.e. when $\zeta\bar\zeta=1/(-\Lambda)$. It can easily be seen that this simply
corresponds to anti-de~Sitter infinity.

Let us finally observe that, just as for the case~2 waves in the Minkowski
background, the above wave surfaces in de~Sitter and anti-de~Sitter backgrounds
are outside the expanding torus or hyperboloid respectively. They therefore
foliate a decreasing portion of the complete background for increasing time
coordinate~$Z_0$.

\section{The de~Sitter and anti-de~Sitter backgrounds for case 1 solutions}

Let us now consider the case 1 solutions in which $\Lambda\ne0$ and $\tau\o$ has
the canonical form (\ref{tauo1}). In this case, the line element (\ref{metric})
in the conformally flat limit (\ref{cflimit}) takes the form  
 $$ \d s^2 =2\d u\left[ \d r
-{(1+\Lambda\zeta\bar\zeta)^2\over(1-\Lambda\zeta\bar\zeta)^2}
\Lambda r^2\d u 
+{4\Lambda\,r\,(\bar\zeta\d\zeta +\zeta\d\bar\zeta)
\over(1-\Lambda\zeta\bar\zeta)(1+\Lambda\zeta\bar\zeta)} \right] 
-{2\,\d\zeta\,\d\bar\zeta\over(1+\Lambda\zeta\bar\zeta)^2}. $$
 Putting \ $r=(Q^2/P^2)v$ \ (with $Q=1-\Lambda\zeta\bar\zeta$), \ 
 this can be written as 
 $$ \d s^2 =2{Q^2\over P^2}\,\big(\d u\,\d v 
-\Lambda\,v^2\,\d u^2 \big)
-{2\over P^2}\,\d\zeta\,\d\bar\zeta. $$ 
 Clearly, this must be (anti-)de~Sitter space which can be represented as
the four-dimensional hyperboloid (\ref{hyp}) embedded in the five-dimensional
Minkowski space (\ref{5dMink}). In this case, the appropriate parametrization is
given by 
 \begin{equation}
 \begin{array}{l}
{\displaystyle Z_0 ={Q\over\sqrt2\,P}(u+v+\Lambda u^2v) } \\[12pt] 
{\displaystyle Z_1 ={Q\over\sqrt2\,P}(u-v+\Lambda u^2v) } \\[12pt] 
{\displaystyle Z_2+iZ_3 =\sqrt2{\zeta\over P} } \\[12pt]  
{\displaystyle Z_4 =a{Q\over P} (1+2\Lambda uv)} 
 \end{array} \qquad \Leftrightarrow \qquad
 \begin{array}{l} 
{\displaystyle 
 u= \sqrt2\,a {\varepsilon(Z_4-\sqrt{-\varepsilon{Z_0}^2
+\varepsilon{Z_1}^2+{Z_4}^2})\over Z_0-Z_1} }
\\[12pt]  {\displaystyle 
 v= {a\over\sqrt2}{Z_0-Z_1\over\sqrt{-\varepsilon{Z_0}^2
+\varepsilon{Z_1}^2+{Z_4}^2}} } \\[18pt] 
{\displaystyle 
 \zeta= {\sqrt2\,a(Z_2+iZ_3)\over 
a+\sqrt{a^2-\varepsilon{Z_2}^2-\varepsilon{Z_3}^2}} } 
 \end{array} 
 \end{equation}

For the case in which $\Lambda>0$ ($\varepsilon=1$), it can be seen that the wave
surfaces, which are given by $u=u_0$, are identical to those for case~2 with
$\Lambda>0$, but with the roles of $Z_2$ and $Z_4$ interchanged. (For this
de~Sitter background,
$Z_2$ and $Z_4$ are both spacelike coordinates and their interchange is trivial.)
This result is consistent with the fact that these two cases are equivalent for a
positive cosmological constant. The wave surfaces are therefore a family of
hemispheres of common constant area $4\pi a^2$ which are tangent to the expanding
torus 
 $$ {Z_2}^2+{Z_3}^2= a^2 \qquad {\rm and} \qquad 
{Z_1}^2+{Z_4}^2={Z_0}^2, $$ 
 which here corresponds to the coordinate singularity \
$1-\Lambda\zeta\bar\zeta=0$.

The case 1 space-times, however, are generalisations of the {\sl pp}-waves in the
sense that they reduce to the {\sl pp}-waves as $\Lambda\to0$. In this case, for
$\Lambda>0$ the background universe is closed, and the analogue of the plane
wave surfaces which occur in a Minkowski background are spherical wave surfaces
of equal constant area~$4\pi a^2$. These spherical surfaces must clearly
intersect each other. So it is not surprising that it is appropriate to restrict
them to hemispheres, although it is perhaps unexpected that this leads to the
same foliation of the space-time as that for case~2.

\goodbreak
Now consider the case for which $\Lambda<0$. In this case, the wave surfaces
$u=u_0$ are now given in this five-dimensional representation by the intersection
of the hyperboloid (\ref{hyp}) with $\varepsilon=-1$ with the hyperplane 
 $$ \textstyle{ \big(1-{u_0^2\over2a^2}\big)Z_0
+\big(1+{u_0^2\over2a^2}\big)Z_1 -{\sqrt2\,u_0\over a}\,Z_4 =0. } $$ 
 Putting 
 $$ \sin\alpha={-{\sqrt2\,u_0\over a} \over1+{u_0^2\over2a^2}}, \qquad 
\cos\alpha={1-{u_0^2\over2a^2}\over1+{u_0^2\over2a^2}}, $$ 
 this becomes 
 $$ Z_1+\cos\alpha\,Z_0+\sin\alpha\,Z_4 =0. $$ 
 This corresponds to a family of planes which are rotated relative to each
other in the section of the timelike coordinates ($Z_0,Z_4$). These planes
intersect the four-hyperboloid in a way which parametrizes the complete
space-time.

In this case, there is no singularity as the term \ $a^2+{Z_2}^2+{Z_3}^2$
\ cannot be zero. The apparent coordinate singularity that occurs in the
metric when \ $P=1+\Lambda\zeta\bar\zeta=0$ \ again corresponds to anti-de~Sitter
infinity. These space-times are generalisations of the {\sl pp}-waves and, for
$\Lambda<0$, the background universe is open and the analogue of the plane wave
surfaces are hyperboloids which foliate the entire universe.

\section{The anti-de~Sitter background for case 3 solutions}

Finally, let us consider the case 3 solutions in which $\Lambda$ is necessarily
negative, $k=0$ and $\tau\o$ has the canonical form (\ref{tauo3}). In this case,
the line element (\ref{metric}) in the conformally flat limit (\ref{cflimit}) is 
 $$ \d s^2 =2\d u\big(\d r +W\d\zeta +\bar W\d\bar\zeta\big) 
-2P^{-2}\d\zeta\d\bar\zeta, $$ 
 where \ $W=2(\bar\tau\o/P)\,r$. \ Putting  \ $r=(Q^2/P^2)v$ \ where \
$Q=(1+\sqrt{-\Lambda}\,\zeta)(1+\sqrt{-\Lambda}\,\bar\zeta)$, \ this becomes 
 $$ \d s^2=2{Q^2\over P^2}\,\d u\,\d v -{2\d\zeta\,\d\bar\zeta\over P^2}. $$ 
 This must be the anti-de~Sitter space which can be represented as a
four-dimensional hyperboloid (\ref{hyp}) with $\varepsilon=-1$ embedded in a
five-dimensional Minkowski space-time (\ref{5dMink}).  In this case, the
parametrization can be expressed in the form 
 \begin{equation}
 \begin{array}{l}
{\displaystyle Z_0 ={(u+v)\over\sqrt2}{Q\over P}} \\[12pt] 
{\displaystyle Z_1 ={(u-v)\over\sqrt2}{Q\over P} } \\[12pt] 
{\displaystyle Z_2 = {(\zeta+\bar\zeta)\over\sqrt2\>P} 
+{uv\over a}{Q\over P} } \\[12pt] 
{\displaystyle Z_3 =-{i(\zeta-\bar\zeta)\over\sqrt2\>P} } \\[12pt] 
{\displaystyle Z_4 =\left( a-{uv\over a} \right){Q\over P}
 -{(\zeta+\bar\zeta)\over\sqrt2\>P} }
 \end{array} \qquad \Leftrightarrow \qquad
 \begin{array}{l}
{\displaystyle u= {a\over\sqrt2}{(Z_0+Z_1)\over(Z_2+Z_4)} } \\[24pt] 
{\displaystyle v= {a\over\sqrt2}{(Z_0-Z_1)\over(Z_2+Z_4)} } \\[24pt] 
{\displaystyle  \zeta= \sqrt2\,a \left[{(Z_2+iZ_3+Z_4)^2-a^2 \over
(a+Z_2+Z_4)^2+{Z_3}^2 }\right] } 
 \end{array} 
 \end{equation}

The wave surfaces \ $u=$~const. \ are located on the intersection of the
hyperboloid with the null hyperplane 
 $$ Z_0+Z_1={\sqrt2u_0\over a}(Z_2+Z_4). $$ 
 The complete family of these hyperboloidal wave surfaces foliate the entire
background space-time and, again, the apparent singularity at $P=0$ simply
corresponds to anti-de~Sitter infinity.

The type N cases of these solutions have been described in detail by Siklos
\cite{Siklos85} and by Podolsk\'y \cite{Podol98}, \cite{Podol01} in different
coordinate systems.

\section{Conclusions}

We have presented the complete family of type~III solutions of Kundt's class with
a non-zero cosmological constant. These are new and contain all the previously
known special cases. We have also found a more general type~N solution that had
previously been overlooked. A classification of all these space-times has been
presented in terms of the signs of the cosmological constant and the parameter
$k$. These give rise to different canonical forms for the
spin-coefficient~$\tau$. We have described the physical interpretation of these
solutions in terms of the global geometry of the family of wave surfaces, at
least in the weak field limit, and we have argued that the weak singularities
which arise in the space-times can be interpreted in terms of the caustics formed
as the envelopes of wave surfaces.

\section*{Acknowledgements}

We are grateful to Pavel Krtou\v{s} of Charles University in Prague for helpful
discussions on these topics. JBG is grateful for hospitality at Charles
University. This and the work of JP was partly supported by the grants
GA\v{C}R-202/02/0735 and GAUK~166/2003.

\newpage
\section*{Appendix}

With the assumptions introduced in section~2, namely \
$\kappa=\sigma=\rho=\epsilon=\lambda=0$, \ $\alpha=\bar\tau-\bar\beta$, \ 
$\pi=-\bar\tau$ \ and the only nonzero components of the curvature tensor being
$\Psi_3$, $\Psi_4$, $\Phi_{22}$ and $\Lambda$, the Ricci identities become 
 \begin{eqnarray}
 D\tau &=&0 \nonumber \\
 D\beta &=&0 \nonumber \\ 
 D\gamma &=& -\tau\bar\tau -\Lambda \nonumber \\ 
 \delta\tau &=&2\tau\beta \nonumber \\
 D\mu &=& 0 \nonumber \\
 D\nu+\Delta\bar\tau &=& -\bar\tau(\gamma-\bar\gamma) +\Psi_3  \nonumber \\
 \bar\delta\nu &=& -\nu(\bar\tau-2\bar\beta) +\Psi_4 \nonumber \\
 \bar\delta\beta+\delta\bar\beta &=&\tau\bar\tau +\tau\bar\beta +\bar\tau\beta 
-4\beta\bar\beta +\Lambda \nonumber \\
 \bar\delta\mu &=& -\bar\tau\bar\mu  +\Psi_3  \nonumber \\ 
 \delta\nu-\Delta\mu &=&\mu^2 +\mu(\gamma+\bar\gamma)
+\bar\nu\bar\tau -2\beta\nu +\Phi_{22} \nonumber \\
 \delta\gamma-\Delta\beta &=&\mu\tau -\beta(\gamma-\bar\gamma-\mu) \nonumber \\ 
 \bar\delta\tau &=&2\tau(\bar\tau-\bar\beta) +2\Lambda \nonumber \\
 \Delta\bar\tau-\Delta\bar\beta-\bar\delta\gamma &=& 
(\bar\tau-\bar\beta)(\bar\gamma-\bar\mu) +\gamma(\bar\beta-\bar\tau) 
-\Psi_3 \nonumber 
 \end{eqnarray}
 For the tetrad (\ref{tetrad}) and the line element (\ref{metric}), the metric
equations are 
 \begin{eqnarray}
 DH &=&\gamma+\bar\gamma \nonumber \\
 D(P\bar W) &=& 2\tau \nonumber \\ 
 \delta H-\Delta(P\bar W) &=&\bar\nu 
 +(\mu-\gamma+\bar\gamma)P\bar W \nonumber \\
 \delta(PW)-\bar\delta(P\bar W) &=& \bar\mu-\mu +(\tau-2\beta)PW
-(\bar\tau-2\bar\beta)P\bar W \nonumber \\
 DP&=&0 \nonumber \\
 \Delta P &=& (\mu-\gamma+\bar\gamma)P \nonumber \\
 \delta P &=& (\tau - 2\beta)P \nonumber 
 \end{eqnarray}
 These, with the radial equations above, imply that the $P$ is independent of $r$
and the other metric functions have the structures 
 \begin{eqnarray}
 H &=& -(\tau\o\bar\tau\o+\Lambda)r^2 +2G\o r +H\o \nonumber \\
 W &=& {2\bar\tau\o\over P}\,r +W\o \nonumber 
 \end{eqnarray}
 With \ $\alpha=\bar\tau-\bar\beta$ \ and \ $\pi=-\bar\tau$, \ the other
non-zero spin coefficients are thus given by 
 \begin{eqnarray}
 \tau &=& \tau\o \nonumber \\
 \beta &=& {\textstyle{1\over2}}(\tau\o-P_{,\bar\zeta}) \nonumber \\
 \gamma &=& -(\tau\o\bar\tau\o+\Lambda)r+G\o 
+{\textstyle{1\over2}}im\o \nonumber \\
 \mu &=& im\o -(\log P)_{,u} \nonumber \\
 \nu &=& 2\Big[ PG\o_{,\zeta} +(\tau\o\bar\tau\o+\Lambda)PW\o
-\bar\tau\o_{,u} +\bar\tau\o(\log P)_{,u} \Big]r \nonumber\\
   && \qquad +PH\o_{,\zeta} +2\bar\tau\o H\o -2PG\o W\o -PW\o_{,u} \nonumber 
 \end{eqnarray} 
 where 
 $$ im\o= -{\textstyle{1\over2}} P \big[ W\o_{,\bar\zeta} -\bar W\o_{,\zeta} \big]
-P(\tau\o W\o-\bar\tau\o\bar W\o). $$ 
 The remaining Ricci identities still need to be satisfied.

\end{document}